\documentclass[12pt,a4paper]{article}
\usepackage{graphicx}
\usepackage{times}
\def\sun{\hbox{$\odot$}}
\title{\bf Radio observations of massive stars}
\author{Ronny Blomme\\
\vspace{1cm}\\
\normalsize Royal Observatory of Belgium, Ringlaan 3, B-1180 Brussel, Belgium \\ 
}
\date{\mbox{}}
\begin{document}
\maketitle
\pagestyle{empty}
%
%
\def\bull{\vrule height .9ex width .8ex depth -.1ex}
\makeatletter
\def\ps@plain{\let\@mkboth\gobbletwo
\def\@oddhead{}\def\@oddfoot{\hfil\tiny\bull\quad
``The multi-wavelength view of hot, massive stars''; 39$^{\rm th}$ Li\`ege Int.\ Astroph.\ Coll., 12-16 July 2010 \quad\bull}%
\def\@evenhead{}\let\@evenfoot\@oddfoot}
\makeatother
%
%
\def\beginrefer{\section*{References}%
\begin{quotation}\mbox{}\par}
\def\refer#1\par{{\setlength{\parindent}{-\leftmargin}\indent#1\par}}
\def\endrefer{\end{quotation}}
%
%
{\noindent\small{\bf Abstract:} 
Detectable radio emission occurs during almost 
all phases of massive star evolution.
I will concentrate on the thermal and non-thermal continuum emission from 
early-type stars. The thermal radio emission is due to free-free interactions 
in the ionized stellar wind material. Early ideas that this would lead to an 
easy and straightforward way of measuring the mass-loss rates were thwarted
by the presence of clumping in the stellar wind. Multi-wavelength 
observations provide important constraints on this clumping, but do not allow 
its full determination.

Non-thermal radio emission is associated with binarity. This conclusion was 
already known for some time for Wolf-Rayet stars and in recent years it has 
become clear that it is also true for O-type stars. 
In a massive-star binary, the 
two stellar winds collide and around the shocks a fraction of the electrons 
are accelerated to relativistic speeds. Spiralling in the magnetic field 
these electrons emit synchrotron radiation, which we detect as non-thermal 
radio emission. The many parameters that influence the resulting non-thermal 
radio fluxes make the modelling of these systems particularly challenging, 
but their study will provide interesting new insight into massive stars.
}
%
%
\section{Introduction}
\label{sect introduction}

Radio emission in massive stars occurs in almost all stages of stellar 
evolution. 
It starts with
the molecular clouds in which the stars are formed: these can
be detected in the radio through their maser emission (Elitzur 1992, Chapt.~8;
Van der Walt, these proceedings).
As the H II region forms, we can detect it in its atomic recombination
lines, as well as in its thermal and non-thermal continuum emission
(Rohlfs \& Wilson 2000, Chapt.~10 \& 13).
In the red supergiant phase, molecular
maser emission again dominates (Elitzur 1992, Chapt.~8). 
During the Luminous Blue Variable (LBV) phase, 
the ionized material in the nebula emits
at radio wavelengths (Umana et al. 2010; Umana, these proceedings).
At the end-phase of evolution, compact objects are
responsible for the radio emission in pulsars, massive X-ray binaries
and micro-quasars (Paredes 2009). Various interactions with the 
interstellar medium, such as bowshocks, bubbles and supernova remnants
are also detectable in radio emission (Rohlfs \& Wilson 2000, Chapt.~10;
Cappa et al. 2002).
The present review, however, will concentrate on the main-sequence 
evolutionary phase, 
as well as on the blue supergiant and Wolf-Rayet (WR) phases, where
the radio emission 
is due to the thermal and non-thermal processes in stellar winds.

\subsection{History}

Historically, the first massive star detected at radio wavelengths
was the LBV star P Cygni. Wendker et al.~(1973)
found fluxes of 
$9 \pm 2$ mJy\footnote{1 mJy = $10^{-29}$ W m$^{-2}$ Hz$^{-1}$ 
                             = $10^{-26}$ erg cm$^{-2}$ Hz$^{-1}$ s$^{-1}$.}
at 6 cm and $15 \pm 3$ mJy at 3 cm.
Later, Wendker et al.~(1975) were the first to detect a Wolf-Rayet star,
WR 136, while
studying its ejecta nebula NGC~6888. 
The first radio-detected O-type star was, not surprisingly,
$\zeta$~Pup (Morton \& Wright 1978).

In 1981, the 27-telescope Very Large Array 
(VLA\footnote{{\tt http://www.vla.nrao.edu/}})
became available, 
which presented a major step forward compared with the single-dish
telescopes that had been used till that time. The 27 antennas
not only provide a larger collecting surface area, but they are 
furthermore used as a radio interferometer. In such an interferometer,
the spatial resolution is determined by the largest distance between
two antennas. The positions of the VLA antennas are changed approximately
every trimester, allowing astronomers to study the radio sky with
different spatial resolutions. The high sensitivity of the VLA
allowed Bieging et al.~(1982) to make a survey of Wolf-Rayet stars
(they detected 8 of the 13 stars studied). Later, Bieging et al.~(1989)
also surveyed OB stars, detecting 18 of the 88 stars studied.

\subsection{Theory}

In parallel, there were also theoretical developments in understanding
and modelling the origin of the radio emission from these hot, massive
stars. Within a very short time span various authors deduced that the
emission is due to free-free processes in the ionized material of the
stellar winds (Seaquist \& Gregory 1973; Olnon 1975; Panagia \& Felli 1975;
Wright \& Barlow 1975).  Working out the details, 
a simple equation is found that relates the radio
flux ($S_\nu$) to the mass-loss rate ($\dot{M}$, in $M_{\sun}$ 
yr$^{-1}$):
\begin{equation}
S_\nu = 2.24 \times 10^{11} \frac{1}{D^2} 
        \left(\frac{\dot{M}}{\mu v_{\infty}}\right)^{4/3}
        \left(\frac{\gamma g Z^2}{\lambda}\right)^{2/3} \quad {[\rm mJy]},
\label{eq smooth wind}
\end{equation}
where $D$ is the distance to the star (in kpc), $v_{\infty}$ is the terminal 
wind velocity (km s$^{-1}$), $\lambda$ is the radio wavelength (cm)
and $g$ is the Gaunt factor.
The chemical composition of the wind is encoded in the $\mu$ factor, and
the ionization state in the $\gamma Z^2$ factors.
Eq.~\ref{eq smooth wind} suggests that radio observations are an easy
way to determine mass-loss rates: we do not need to know the detailed
ionization balance of some trace species (as we need for the 
analysis of ultraviolet P Cygni profiles),
but only the gross ionization properties of the wind. We also do not need 
to know the shape of the velocity law (as we do need for H$\alpha$, infrared
and millimetre work), but only the terminal velocity.

The radio flux shows a near-power law dependence on the wavelength. Taking
into account the weak wavelength dependence of the Gaunt factor, we have
that 
$S_\nu \propto 1/\lambda^{+0.6}$, where the exponent $\alpha = +0.6$ 
is called the spectral
index. It is also important to realize that the geometric region from which we 
receive the radio emission is quite extended. For an O-type star such 
as $\zeta$~Pup, the 6-cm formation region is beyond $\sim 100 R_*$. Furthermore,
this region depends on the wavelength, because
the optical depth for free-free absorption is proportional to $\lambda^2$.
At longer wavelengths, the formation region is formed further away from
the star.

\subsection{Non-thermal emission}
\label{sect introduction NT}

Although the radio emission for most early-type
stars is due to free-free emission,
it became clear already early on that a number of anomalies exist.
Abbott et al.~(1980) found that the radio mass-loss rate for the 
O4~V star 9~Sgr
was a factor 40 higher than that derived from the ultraviolet or H$\alpha$
line profiles.
Such a difference is much higher than can be accounted for by the
intrinsic errors.
From spatially resolved observations, it is also possible to derive
a brightness temperature (or a lower limit, if the source is
unresolved).
The brightness temperature of Cyg~OB2 No.~9 was found by 
White \& Becker~(1983) to be higher than 300,000 K. It is
very unlikely, however,
that the electrons and ions responsible for the free-free
emission would be at so high a temperature. Abbott et al.~(1984)
found flux variability in 9~Sgr and Cyg~OB2 No.~9 which 
was too high to
be attributed to changes in the stellar wind parameters. They
also measured radio flux values that deviate significantly from the 
+0.6 spectral index, with values of $\alpha \approx 0.0$,
or even negative in later observations (Bieging et al.~1989). 
All this indicates
that, besides the free-free emission, a second mechanism is operating in these 
stars, which we call non-thermal emission.

\section{Thermal radio emitters}

Eq.~\ref{eq smooth wind} has been used to determine the mass-loss rates of
hot, massive stars. One example is the large set of Wolf-Rayet
mass-loss rate determinations by Cappa et al.~(2004). 
The equation has
also been used to study the mass-loss rate across the bi-stability jump
(Benaglia et al.~2008). Theoretical modelling predicts a change in ionization 
around 21,000 K that substantially changes the mix of lines responsible for the
radiative driving of the wind. The radio observations indeed indicate that
around the bi-stability jump, stellar winds have a higher efficiency 
than expected from the general declining trend towards later spectral
types.

\subsection{Clumping}

The major problem with Eq.~\ref{eq smooth wind}, however, is that it does
not include the effect of clumping or porosity. Many indicators show the
presence of substantial clumping in stellar winds:
the Phosphorus V discrepancy, 
X-ray spectroscopy,
the electron scattering wings of WR emission lines, 
the presence of subpeaks on WR emission lines that are seen moving outward, ...
(Crowther 2007; Puls et al. 2008).
Also from a theoretical point of view, considerable clumping is expected
due to the instability of the radiation driving mechanism of the stellar
wind (Owocki \& Rybicki 1984).

Because free-free emission is a process that depends on the density-squared,
it is influenced by this clumping.
Eq.~\ref{eq smooth wind} can be extended to include some simple model
of clumping that assumes all the wind
material is concentrated in clumps, with no interclump material. All clumps 
have the same clumping factor given by $f_{\rm cl} = <\rho^2>/<\rho>^2$,
where the angle brackets indicate an average over the volume in which
the radio continuum at that wavelength is formed. 
The resulting equation is (Abbott et al. 1981):
\begin{equation}
S_\nu = 2.24 \times 10^{11} \frac{1}{D^2} 
        \left(\frac{\dot{M} \sqrt{f_{\rm cl}}}{\mu v_{\infty}}\right)^{4/3}
        \left(\frac{\gamma g Z^2}{\lambda}\right)^{2/3}.
\label{eq clumped wind}
\end{equation}
Sometimes, in the literature,
a volume filling factor $f$ is used. With our assumption of no
interclump material, there is a direct relation between $f$ and the
clumping factor: $f=1/f_{\rm cl}$. In the above equation we have
also assumed that the clumps are optically thin; optically thick clumps
would result in porosity (Owocki et al.~2004).

Eq.~\ref{eq clumped wind} considerably changes the interpretation of an
observed radio flux. All we can derive from the observations is the
combined quantity $\dot{M} \sqrt{f_{\rm cl}}$.
A given flux can thus correspond to a certain mass-loss
rate assuming a smooth wind ($f_{\rm cl} = 1$), or to a lower mass-loss
rate assuming a clumped wind ($f_{\rm cl} > 1$).
This uncertainty in the mass-loss rates has important consequences
for stellar evolution: if the rates during the main-sequence evolution
are too low, mass-loss episodes during other evolutionary phases need
to be invoked to explain the existence of 
Wolf-Rayet stars (Smith \& Owocki 2006).

\begin{figure}[h]
\centering
\includegraphics[width=14cm]{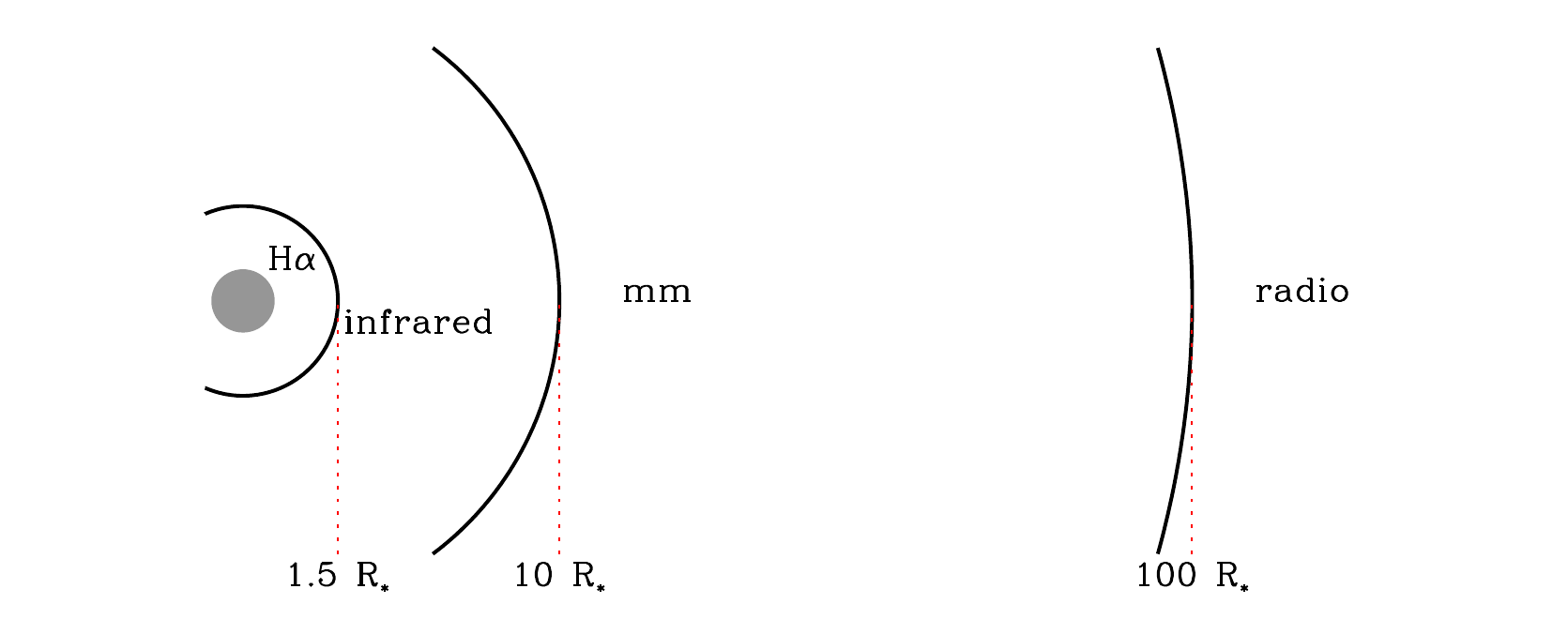}
\caption{The formation regions of various density-squared observational
indicators in the stellar wind of a typical O-type star.}
\label{fig formation regions}
\end{figure}

\subsection{Clumping gradients in OB stellar winds}

From radio data alone, it is difficult to make a distinction between 
smooth and clumped winds. But a number of other observational indicators
are also sensitive to clumping because they depend on density-squared
processes (as does the radio emission). These indicators include the 
H$\alpha$ spectral line, the infrared and the millimetre continuum.
Interestingly, these different indicators are formed in different regions
of the wind (Fig.~\ref{fig formation regions}). 
Very approximate values for an O-type star with a 
strong stellar wind are: H$\alpha$ within
$1.5$\,$R_*$ of star, infrared up to a few $R_*$, millimetre from about
$10$\,$R_*$ and radio from $\sim 100$\,$R_*$.

This opens up the possibility to compare the clumping in different regions
of the wind, and to see if there is a gradient in the clumping. 
Lamers \& Leitherer (1993) were the first to attempt this, by comparing the 
mass-loss rate derived from the H$\alpha$ line with that from the radio
(assuming a smooth wind).
They found no significant differences and therefore concluded that 
significant clumping in stellar winds was unlikely. 

Runacres \& Blomme (1996) compared the infrared, millimetre and radio fluxes
with smooth wind models for a sample of 18 OB stars. The models were fitted
through the observed visual and near-infrared continuum fluxes, which
fixes the distance and interstellar extinction. They determined the
mass-loss rate
so that a good fit to the observed radio fluxes was also obtained.
They then checked how well this smooth wind model agreed with the
observed far-infrared and millimetre fluxes (which were not used in the 
fitting procedure). Four stars 
($\alpha$~Cam, $\delta$~Ori~A,
$\kappa$~Ori and $\zeta$~Pup) show fluxes that are significantly higher
than the smooth wind model. 

\begin{figure}[h]
\centering
\includegraphics[width=16cm]{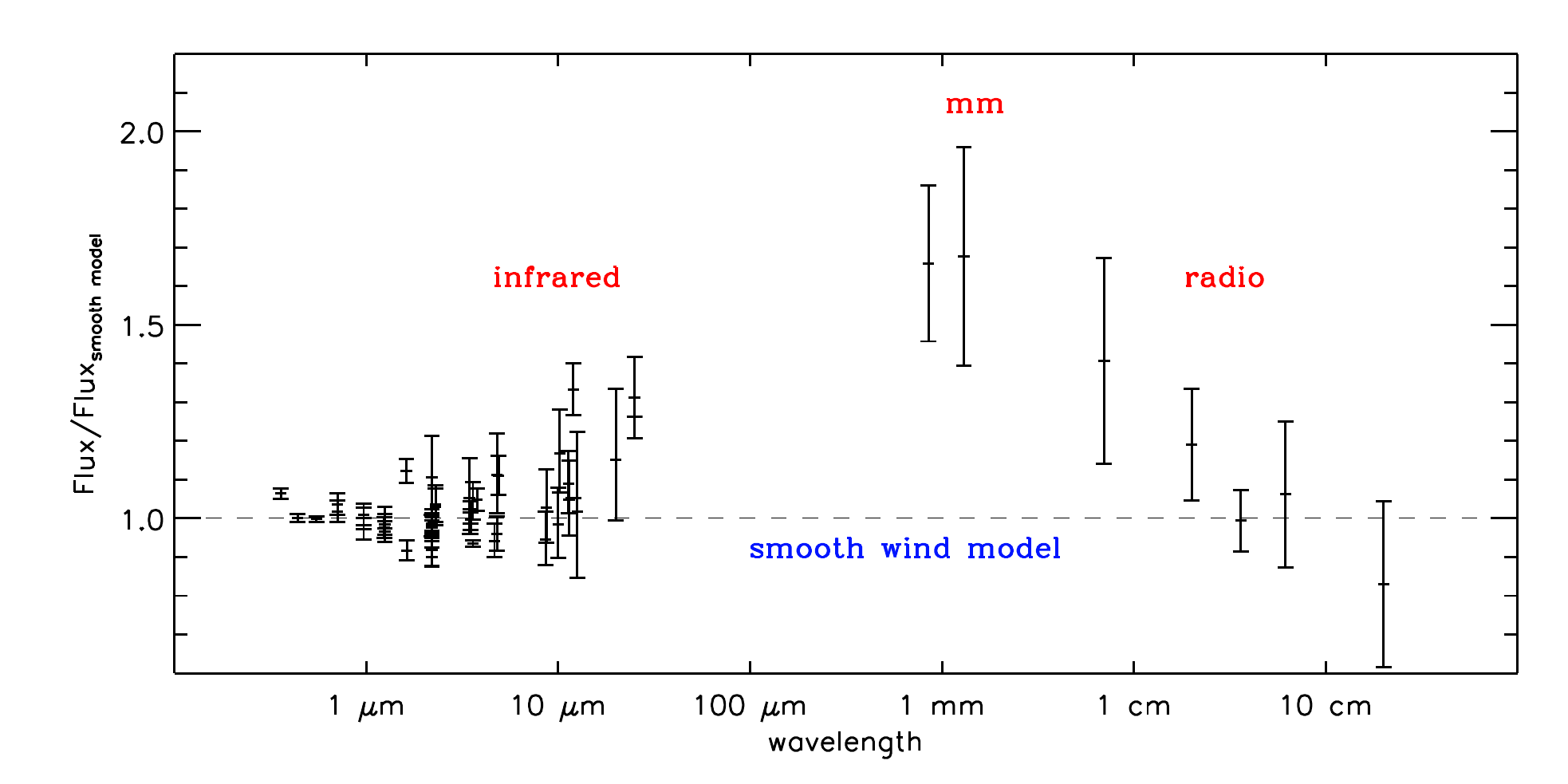}
\caption{The $\epsilon$~Ori fluxes from visual to radio wavelengths, 
compared with a smooth wind model. The flux
excess in the millimetre region is interpreted as indicating a higher clumping
factor in the millimetre formation region. Based on Blomme et al.~(2002), with
some new radio data added.}
\label{fig eps ori}
\end{figure}

Later, an even better example was found: for the
bright B0 Ia star $\epsilon$~Ori, Blomme et al.~(2002) combined new data
with archival observations and applied the same technique. 
Fig.~\ref{fig eps ori} shows the observed fluxes as a function of
wavelength, covering the range from visual to radio wavelengths.
The y-axis shows the observed flux divided by the smooth wind flux. Any 
significant excess above the y=1 line is interpreted as being due to
clumping in the wind. A clear excess is seen at millimetre wavelengths,
with a possible onset already in the far-infrared and a continuation
in the radio region. This indicates the presence of more clumping in the
geometric region where the millimetre flux is formed.

It is important
to realize that in this type of work we need to assume that the wind region
where the radio emission is produced, 
is not clumped, otherwise we would just introduce another
parameter on which we have no observational constraints. But it is of
course quite likely that the radio formation region is also clumped.
The downward gradient seen in the 7~mm to 20~cm
radio fluxes (Fig.~\ref{fig eps ori}) is 
suggestive of some decrease of clumping in the radio formation region,
as we go further out in the wind.

Puls et al.~(2006) not only studied the continuum fluxes, but also included
the H$\alpha$ line, for a sample of 19 O-type supergiants and giants. The 
first thing they noticed is that the radio mass-loss rates (assuming a
smooth wind) are in better agreement with the predicted wind-momentum
luminosity relation (Vink et al. 2000) than the H$\alpha$ mass-loss
rates (again, assuming a smooth wind). A closer inspection reveals that
it is mainly the stars with H$\alpha$ in emission (i.e. those with
a stronger stellar wind) that are most discrepant.

Puls et al.~(2006) then introduced a clumped wind model. The wind was divided
into 5 regions, and in each region a constant clumping factor was taken.
These clumping factors were then adjusted to obtain a good fit to the
observed H$\alpha$, infrared, millimetre and radio data. Because of
the reason explained above, the radio region was assumed not to be clumped.
A major result from their work is the behaviour of the clumping factor
in the H$\alpha$ formation region (their Fig.~10). For stars with
low mass-loss rates, the clumping factors are not significantly different
from 1. But for stars with higher mass-loss rates, the H$\alpha$
clumping factors average around 4. This means that the geometric region
where H$\alpha$ is formed (i.e. close to the stellar surface) is more
clumped than the radio formation region, at least for OB stars with
strong winds.

\subsection{Clumping gradients in Wolf-Rayet stellar winds}

Such a gradient in the clumping factor might also exist in Wolf-Rayet stars.
Nugis et al.~(1998) studied the spectral index between the infrared and
radio wavelengths for 37 WR stars. 
According to Eq.~\ref{eq smooth wind}, this spectral index
should be $\alpha = +0.6$ for a wind that has the same clumping everywhere.
However, Nugis et al. measured $\alpha$ values from $+0.66 \pm 0.01$ up to 
$+0.88 \pm 0.04$,
with most spectral indices being larger than $+0.7$ (their Fig.~1). This again
indicates more clumping in the inner part of the wind compared with the outer
part. Nevertheless, the interpretation of these results is more complicated
than for OB stars, because the ionization in a Wolf-Rayet wind can change
sufficiently to influence the spectral index. Furthermore, for at least
one WR star, Montes et al. (these proceedings) ascribe the higher than $+0.6$
spectral index to thermal emission in the increased density region of a 
colliding-wind binary.

In summary, it is clear that OB stars with strong winds show a gradient
in their stellar wind clumping, with stronger clumping in the inner part
of the wind. Puls et al.~(2006) also point out that these results
are not in agreement with the hydrodynamical models of 
Runacres \& Owocki (2005), 
which predict more clumping in the radio than in H$\alpha$.
It is possible however that even minor changes in the input physics
of these hydrodynamical models might change this conclusion. A similar
clumping gradient probably also exists in the winds of Wolf-Rayet stars,
with the important caveat that ionization changes also play a role which
are not easy to distinguish from clumping.

\section{Non-thermal radio emitters}
\label{sect non-thermal}

In Sect.~\ref{sect introduction NT}, we saw that some hot, massive stars
also show non-thermal radio emission. This emission is characterized by
a spectral index significantly smaller than the $+0.6$ value for thermal
emission and by a high brightness temperature. It is also frequently associated
with variability in the radio fluxes, though this characteristic in itself
is not sufficient to identify non-thermal emission: e.g., the LBV P~Cygni
shows radio variability, but this is due to ionization changes in the wind,
not to non-thermal emission (Exter et al. 2002).

In the literature, various values for the incidence of non-thermal emission
are cited. Abbott et al.~(1986) claim that 12 \% of the Wolf-Rayet stars
are non-thermal and Bieging et al.~(1989) give at least 24 \% of the OB stars.
Later, much higher numbers were claimed: 20 -- 30 \% of the detected WR stars 
(Cappa et al. 2004); more than 40 \% of the Wolf-Rayet 
stars (Leitherer et al.~1997; Chapman et al.~1999);
up to 50 \% of detected O-type stars (Benaglia et al. 2001).
The reason for these differences is an observational bias: non-thermal
emitters are intrinsically radio-brighter than thermal ones and therefore 
easier to detect. The higher percentages indicate the probability of
finding a non-thermal emitter in a random sample of massive stars. The
lower percentages by Abbott et al.~(1986) and Bieging et al.~(1989) 
refer to volume-limited samples, and therefore give a better indication
of the incidence of non-thermal emission in massive stars.

\subsection{Binary connection}
\label{sect binary connection}

Quite early on, it was clear that there was some connection between 
non-thermal emission and binarity. Moran et al.~(1989)
used the MERLIN radio telescopes to
spatially resolve the emission from WR 147. They found a thermal
radio emitting source at the position of the Wolf-Rayet star itself
and also a well separated non-thermal source 
between the WR star and a close-by B-type star (this companion
was discovered later in the infrared by Williams et al.~1997). The geometry of
the situation suggests that the non-thermal region is positioned at
the collision region between the wind of the Wolf-Rayet star and that 
of the B-type star.

\begin{figure}[h]
\centering
\includegraphics[width=12cm]{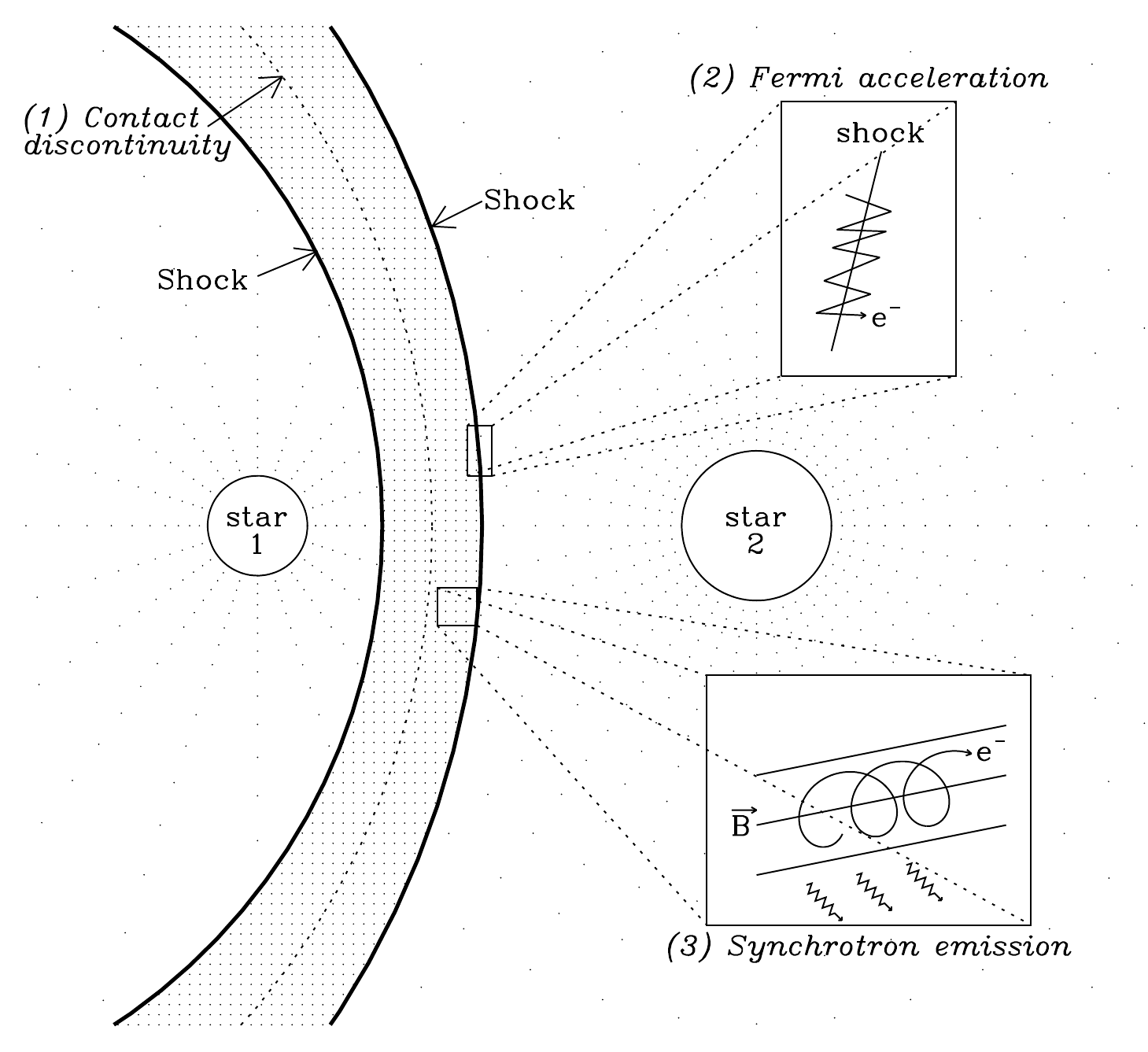}
\caption{Synchrotron emission from colliding winds. {\em (1)} The winds of both 
components collide, creating a shock on either side of the contact
discontinuity. {\em (2)}
At each shock, the Fermi mechanism accelerates a fraction
of the electrons to relativistic speeds. {\em (3)} These relativistic electrons
spiral in the magnetic field and emit synchrotron radiation.}
\label{fig colliding winds}
\end{figure}

A detailed theoretical explanation was provided by Eichler \& Usov (1993).
In a massive binary system, the stellar winds of both components collide
(Fig.~\ref{fig colliding winds}).
This creates a contact discontinuity with a shock on either side. The
position and shape of the contact discontinuity is determined by the
relative strengths (ram pressure) of the two winds: the discontinuity is
closer to the star with the weakest wind, and wraps around that star.

At each shock, a fraction of the electrons is accelerated up to 
relativistic speeds, through the first-order Fermi mechanism
(Bell 1978). In this mechanism, electrons bounce back and forth
across the shock and at each bounce gain some energy from the shock
(Fig.~\ref{fig colliding winds}).
A fraction of the electrons make enough bounces to attain relativistic speeds.
These electrons then spiral around in the magnetic field, emitting 
synchrotron radiation at radio wavelengths. It is this synchrotron radiation
that we see as the non-thermal emission.

The synchrotron emission
has a spectral index of $-0.5$ to $-1.0$
(Bell 1978; Pittard et al. 2006), explaining the
observed spectral index. The emission is due to high-energy particles,
which explains the high brightness temperature. Not all of the synchrotron
emission is detected, however, because part of it (or in some cases, all of it)
is absorbed by the free-free absorption in the stellar winds of both stars.
The effect of the absorption depends on the orbital phase, because the
sightline from the synchrotron emission region to the observer passes
through different parts of the wind(s) as the positions of the stars change
in their orbit. Furthermore, in an eccentric binary, the changing 
separation also generates variability in the intrinsic synchrotron 
emission as the ram pressure of the collision changes. 
The combination of both effects explains the variability
seen in non-thermal radio emitters.

Van der Hucht et al.~(1992) studied a number of Wolf-Rayet stars and 
found a good correlation between binarity and non-thermal emission. 
They also showed
that these binaries are frequently associated with dust emission
(detectable in the infrared) and with X-ray emission. They made the
bold extrapolation {\em ``... that all other non-thermal WR
stars discussed in this paper, as well as all (variable) non-thermal
radio OB stars (...), are actually long-period binaries..."}.

\subsection{Wolf-Rayet binaries}

An outstanding example of a colliding-wind binary is WR~140. It consists
of a WC7 + O4-5 star in a 7.9-yr orbit with a high eccentricity
($e \approx 0.88$). White \& Becker (1995) followed this system
during more than one orbital period and determined the 2, 6 and 20~cm
fluxes approximately every month. The detailed radio light curves
(their Fig.~3) show substantial varibility that is well correlated
with orbital phase and that repeats from one orbit to the next.
The light curves at different wavelengths are reasonably similar to one
another, except that the shorter-wavelength curves 
start rising earlier and decline
more slowly after maximum. The variability is due to the combined
effect of changing synchrotron emission and free-free absorption as
a function of orbital phase. Obtaining a good model fit
to these data is still a considerable challenge (Pittard, these proceedings).

More recently, Dougherty et al.~(2005) have used the Very Long Baseline
Array (VLBA) to spatially resolve the wind-collision region. They obtained
data with a resolution of $\sim 2$~milli-arcsec at 23 epochs,
covering orbital phases 0.74 to 0.97. These data reveal
how the emission region changes orientation and size as a function of orbital
phase. The continued popularity of WR~140 is shown by the many contributions
about this system
at this colloquium (Dougherty et al.; Fahed; Parkin; Russell et al.;
Sugawara et al.; Williams, these proceedings).

Although WR 140 is an outstanding example of the link between binarity
and non-thermal emission, the link cannot be proven with only one, or a
few, examples. A more statistical approach was taken by 
Dougherty \& Williams (2000). From the
23 WR stars they studied, they selected 9 that had a non-thermal spectral 
index, or composite index (between thermal and non-thermal). Seven of these
stars are known binaries (later, Marchenko et al.~(2002) found WR 112 to
have dusty spiral, strongly suggesting it is also a binary). A plot
of the spectral index vs. orbital period (Dougherty \& Williams, their Fig.~1)
shows an interesting correlation:
short-period stars have a nearly-thermal index, while long-period
stars have a non-thermal index. This effect is easily explained with
the colliding-wind model: 
short-period binary components are so close to one another
that all synchrotron emission is absorbed by the free-free absorption 
region in the stellar winds. 
Longer-period binaries have components that are
further away from one another; the synchrotron emitting region is therefore 
largely outside the free-free absorption, and most of the synchrotron emission
gets through to the observer.

\subsection{Single O stars}

While the link between binarity and non-thermal emission is quite strong
for Wolf-Rayet stars,
this was not the case for O-type stars. Some years ago, there were still a
number of non-thermal O stars that were seemingly single.
Before getting carried away by the binary explanation, it is therefore
important to also investigate if single stars can produce non-thermal emission.
After all, there are shocks in the winds of single stars, due to the
instability of the radiative driving mechanism (Owocki \& Rybicki 1984), and
these shocks should be able to accelerate electrons and thus
generate synchrotron emission.

Van Loo et al.~(2006) investigated this possibility for a typical O-star
non-thermal emitter, Cyg~OB2 No.~9. They made a model containing a number
of shocks in the stellar wind. The parameters of these shocks (mainly the
shock jump velocity) were inspired by the Runacres \& Owocki (2005) 
hydrodynamical model. They then calculated the number of relativistic electrons
accelerated at each shock and followed them as they cooled down due to
inverse Compton and adiabatic cooling. From this, the synchrotron emission
was calculated, and -- taking into account the free-free absorption --
the emergent radio flux was determined. There is considerable synchrotron
emission in the inner part of the wind, where the shocks are strong
(high jump velocity). This emission, however, is completely absorbed by
the free-free absorption. In the outer part of the wind, there is also
synchrotron emission, but the shock jump velocity has reduced considerably
there, so there is not enough emission to explain the observations.
The fact that the model cannot explain the Cyg~OB2 No.~9 observations
throws serious doubt on the single-star hypothesis.

\subsection{O star binaries}

More positive proof that Cyg~OB2 No.~9 is a binary was provided by
Van Loo et al.~(2008). Using VLA archive data covering 20 years, they
detected a 2.355-yr period in the radio data (see also Volpi, these
proceedings). The radio fluxes vary with orbital phase and the variations
repeat very well from one orbit to the next. All this indicates that 
Cyg~OB2 No.~9 is a binary. At the same time Naz\'e et al.~(2008) also 
detected binarity in the spectroscopic data. Newer spectroscopic
observations are
presented in Naz\'e et al.~(2010, and these proceedings).

A similar technique was used on HD~168112 observations
by Blomme et al.~(2005). The
amount of data is less than for Cyg~OB2 No.~9, so the result is less
certain. A period between 1 and 2 years is found, with P = 1.4 yr as the
formally best value. There is as yet no spectroscopic confirmation of
the binary status of this star.

As we did for the Wolf-Rayet stars, we need to proceed to a more statistical 
approach to show the link with binarity. Detailed statistics of non-thermal
emission in O-type stars are provided by De Becker (2007) and by
Benaglia (2010). De Becker (his Table~2) shows that many of the 
O-star non-thermal emitters are now confirmed, or at least suspected,
binaries. Although a few stars have not yet been investigated for
multiplicity, we can now confidently confirm the 
Van der Hucht et al. (1992) quote (Sect.~\ref{sect binary connection})
that all non-thermal O-star and Wolf-Rayet star
radio emitters are indeed colliding-wind binaries.

\subsection{Remaining problems}

Although the link between non-thermal emission and binarity is now
quite secure, that does not mean we understand everything about non-thermal
radio emission. One interesting point is that
quite a number of non-thermal emitters are actually multiple
systems.
A nice example is HD 167971, which consists of a 3.3-d eclipsing binary 
and a third light (Leitherer et al. 1987). It is not clear if this 
third component is gravitationally bound to the binary, or whether it is
just a line-of-sight object. Blomme et al.~(2007) studied the VLA
archive data of this system. They did not find any radio variability
on the 3.3-day time scale, as may be expected from such a short-period
binary. But the data show a clear cycle of $\sim 20$~yrs, suggesting
that the third component and the binary are gravitationally bound and
orbit each other with a 20-yr period. Another example of a multiple
system is Cyg~OB2 No.~5 (Kennedy et al. 2010; Dougherty, these proceedings).

\begin{figure}[h]
\centering
\includegraphics[width=17cm]{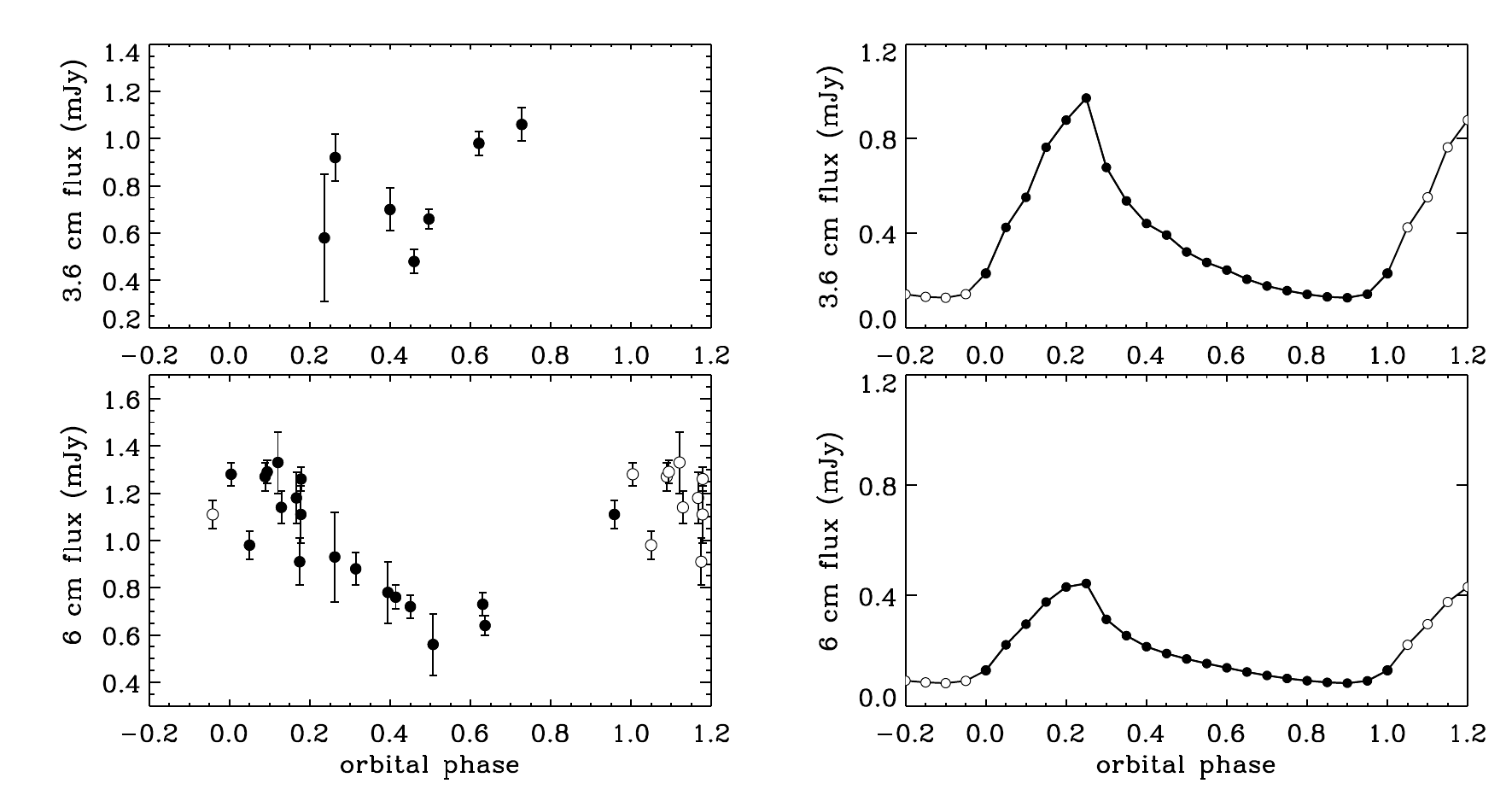}
\caption{Observed (left panel) and calculated (right panel) radio 
fluxes for Cyg OB2 No.~8A. 
The phase range is extended by 0.2 on either side to provide a good overview 
of the variability. Open circles indicate duplication in this extended range.
Phase 0.0 corresponds to periastron.
From Blomme et al.~(2010).}
\label{fig No. 8A}
\end{figure}

A further problem is posed by the short-period binary Cyg~OB2 No.~8A 
(P = 21.9 d, De Becker et al. 2004). 
Although no synchrotron emission was expected from
such a close binary, the VLA data do show orbit-locked variability
(Blomme et al. 2010; see also Fig.~\ref{fig No. 8A}, left panel). This radio
variability is furthermore nearly anti-correlated
to the X-ray variability. Blomme et al.~develop a theoretical model
for this system, starting from an analytical solution for the 
position of the contact
discontinuity. At both shocks, relativistic electrons are injected
and followed as they move away with the flow and as they cool down due to
adiabatic and inverse Compton cooling. The resulting synchrotron emission
is then included in a radiative transfer calculation, together with
free-free absorption, giving the radio flux. By following the two components
along their orbit, an artificial light curve is constructed.
A first model that uses the wind parameters determined by De Becker 
et al.~(2006) fails to explain the observations.
By accepting stellar wind parameters that are different from those expected
for single stars, however, we achieve a better agreement
(Fig.~\ref{fig No. 8A}, right panel). The model can
qualitatively explain many of the observed features of the radio data,
as well as the anti-correlation with the X-ray data. 
Problems remain with explaining the spectral 
index: the model values range between +0.75 and +1.5, while the observed
value remains close to $\sim\;0.0$. Porosity is suggested as a possible 
solution for this problem.

Modelling these colliding-wind binaries can therefore tell us
something about porosity in single-star winds. For this, we need
detailed theoretical models, which is quite challenging work as
the models need to explain not only the radio data, but also the X-ray and
optical spectroscopy observations. A review of modelling colliding-wind
binaries is given by Pittard (these proceedings).

\section{Future radio astronomy}

A number of existing radio facilities are currently being upgraded. 
The VLA is being converted
into the 
EVLA\footnote{{\tt http://science.nrao.edu/evla/}}
(Dougherty \& Perley, these proceedings), resulting in an
order of magnitude improvement in sensitivity and a very large wavelength
coverage. The MERLIN instrument is also being upgraded to 
e-MERLIN\footnote{{\tt http://www.e-merlin.ac.uk/}}.
One of the legacy programmes planned for e-MERLIN is a deep radio survey
of the Cyg OB2 association (PI: R. Prinja; see Willis et al., these
proceedings). As Sect.~\ref{sect non-thermal} has made clear,
Cyg OB2 has provided quite a number of colliding-wind
binaries and these new data are expected to provide more examples, as
well as many detections of thermal radio emission.

On the southern hemisphere, two interferometers are being constructed,
one in South-Africa 
(MeerKAT\footnote{{\tt http://www.ska.ac.za/meerkat/}})
and one in Australia 
(ASKAP\footnote{{\tt http://www.atnf.csiro.au/projects/askap/index.html}}). 
Both
will start with a modest number of telescopes, but MeerKAT will later be
extended to 80 antennas (in 2013-2016), and ASKAP to 36 antennas (2013).
Both instruments will allow much deeper radio observations of the
southern sky than was hitherto possible.

On a longer time scale, the Square Kilometer Array 
(SKA\footnote{{\tt http://www.skatelescope.org/}}) 
will be built,
providing a further order of magnitude improvement in sensitivity
compared with the EVLA. SKA is expected to have 10~\% of its capability
in 2016-2019 with full capability by 2024. Much closer in time is 
the Atacama Large Millimeter Array 
(ALMA\footnote{{\tt http://www.almaobservatory.org/}}), 
consisting of 66 antennas, 
covering the wavelength range 350 $\mu$m -- 9~mm. Early science will
start in 2011, with full operations in 2013.

\section{Conclusions}

The simple relation between thermal radio emission and mass-loss rate
was thought to give us a reliable set of mass-loss rates for hot, massive 
stars. Unfortunately, the presence of clumping in the wind makes this
impossible at the moment. The radio data are very useful however as
a reference point to determine the amount of 
clumping in the inner wind. From this we learned that there is a
clumping gradient in strong stellar winds, with the clumping decreasing
as we move further away from the star. It is important to realize that,
out of necessity, it is assumed there is no clumping in the radio
formation region. It should be realized however that the radio
region is probably clumped as well.

For the non-thermal radio emission, it is now clear that this is
due to colliding-wind binaries, both for Wolf-Rayet and O stars.
The variations in the observed radio light curve are due to a combination of
intrinsically varying synchrotron emission (in an eccentric binary) and
the varying free-free absorption along the sightline to the observer.
The study of these colliding-wind systems is important for a number of
reasons: we can learn more about the Fermi mechanism that accelerates
the electrons, which is also relevant in other astrophysical contexts;
colliding-wind binaries help in the binary frequency determination in clusters;
and they can provide constraints on clumping and porosity in stellar winds.
This last constraint depends on the development of good models for the
observations. Such modelling work is quite challenging, as it has to explain
not only the radio data, but also the X-ray and optical spectroscopy 
observations.

Finally, radio instrumentation is currently being upgraded and new facilities
are planned 
for the near future. For the study of radio emission from hot, massive
stars, this will provide a step forward that is at least as
big as the introduction of the VLA was, some 30 years ago.

%
%
\section*{Acknowledgements}
I would like to thank my colleagues who collaborated with me on this 
interesting subject of radio observation of massive stars, especially
M.~De Becker,
R.~K.~Prinja, 
G.~Rauw,
M.~C.~Runacres, 
J.~Vandekerckhove,
S.~Van Loo and
D.~Volpi.
%
%
\footnotesize
\beginrefer

\refer Abbott, D. C., Bieging, J. H., Churchwell, E., \& Cassinelli, J. P.
       1980, ApJ, 238, 196

\refer Abbott, D. C., Bieging, J. H., \& Churchwell, E. 1981, ApJ, 250, 645

\refer Abbott, D. C., Bieging, J. H., \& Churchwell, E. 1984, ApJ, 280, 671

\refer Abbott, D. C., Bieging, J. H., Churchwell, E., \& Torres, A. V. 
       1986, ApJ, 303, 239

\refer Bell, A. R. 1978, MNRAS, 182, 147

\refer Benaglia, P. 2010, ASP Conf. Proc., 422, 111

\refer Benaglia, P., Cappa, C. E., \& Koribalski, B. S. 2001, A\&A, 372, 952

\refer Benaglia, P., Vink, J. S., Ma\`{i}z Apell\`{a}niz, J.,  et al.
       2008, Rev. Mex. Astron. Astrofis. (Serie de Conferencias), 33, 65

\refer Bieging, J. H., Abbott, D. C., \& Churchwell, E. B. 1982, ApJ, 263, 207

\refer Bieging, J. H., Abbott, D. C., \& Churchwell, E. B. 1989, ApJ, 340, 518

\refer Blomme, R., Prinja, R. K., Runacres, M. C., \& Colley, S. 
       2002, A\&A, 382, 921

\refer Blomme, R., Van Loo, S., De Becker, M., et al. 2005, A\&A, 436, 1033

\refer Blomme, R., De Becker, M., Runacres, M. C., Van Loo, S., \& 
       Setia Gunawan, D. Y. A. 2007, A\&A, 464, 701

\refer Blomme, R., De Becker, M., Volpi, D., \& Rauw, G. 2010, A\&A, 519, A111

\refer Cappa, C. E., Goss, W. M., \& Pineault, S. 2002, AJ, 123, 3348

\refer Cappa, C., Goss, W. M., \& Van der Hucht, K. A. 2004, AJ, 127, 2885

\refer Chapman, J. M., Leitherer, C., Koribalski, B., Bouter, R., \&
       Storey, M. 1999, ApJ, 518, 890

\refer Crowther, P. A. 2007, ARA\&A, 45, 177

\refer De Becker, M. 2007, A\&AR, 14, 171

\refer De Becker, M., Rauw, G., \& Manfroid, J. 2004, A\&A, 424, L39

\refer De Becker, M., Rauw, G., Sana, H., et al. 2006, MNRAS, 371, 1280

\refer Dougherty, S. M., \& Williams, P. M. 2000, MNRAS, 319, 1005

\refer Dougherty, S. M., Beasley, A. J., Claussen, M. J., Zauderer, B. A., \&
       Bolingbroke, N. J. 2005, ApJ, 623, 447

\refer Eichler, D., \& Usov, V. 1993, ApJ, 402, 271

\refer Elitzur, M.  1992, {\em Astronomical Masers}, Astrophysics and space
       science library Vol. 170, Kluwer, the Netherlands

\refer Exter, K. M., Watson, S. K., Barlow, M. J., \& Davis, R. J. 
       2002, MNRAS, 333, 715

\refer Kennedy, M., Dougherty, S. M., Fink, A., \& Williams, P. M.
       2010, ApJ, 709, 632

\refer Lamers, H. J. G. L. M., \& Leitherer, C. 1993, ApJ, 412, 771

\refer Leitherer, C,, Forbes, D., Gilmore, A. C., et al. 1987, A\&A, 185, 121

\refer Leitherer, C., Chapman, J. M., \& Koribalski, B. 1997, ApJ, 481, 898

\refer Marchenko, S. V., Moffat, A. F. J., Vacca, W. D., C\^{o}t\'e, S., \&
       Doyon, R. 2002, ApJ, 565, L59

\refer Moran, J. P., Davis, R. J., Spencer, R. E., Bode, M. F., \&
       Taylor, A. R. 1989, Nature, 340, 449

\refer Morton, D. C., \& Wright, A. E. 1978, MNRAS, 182, P47

\refer Naz\'e, Y., De Becker, M., Rauw, G., \& Barbieri, C.
       2008, A\&A, 483, 543

\refer Naz\'e, Y., Damerdji, Y., Rauw, G., et al. 2010, ApJ, 719, 634

\refer Nugis, T., Crowther, P. A., \& Willis, A. J. 1998, A\&A, 333, 956

\refer Olnon, F. M. 1975, A\&A, 39, 217

\refer Owocki, S. P., \& Rybicki, G. B. 1984, ApJ, 284, 337

\refer Owocki S.P., Gayley K.G., \& Shaviv N.J. 2004, ApJ, 616, 525

\refer Panagia, N., \& Felli, M. 1975, A\&A, 39, 1

\refer Paredes, J. M. 2009, ASP Conf. Proc., 407, 289

\refer Pittard, J. M., Dougherty, S. M., Coker, R. F., O'Connor, E., \&
       Bolingbroke, N. J. 2006, A\&A, 446, 1001

\refer Puls, J., Markova, N., Scuderi, S., et al. 2006, A\&A, 454, 625

\refer Puls, J., Vink, J. S., \& Najarro, F. 2008, A\&AR, 16, 209

\refer Rohlfs, K. \& Wilson, T. L. 2000, {\em Tools of radio astronomy},
       3rd ed., Astronomy and Astrophysics Library, Springer-Verlag,
       Berlin, Heidelberg, New York

\refer Runacres, M. C., \& Blomme, R. 1996, A\&A, 309, 544

\refer Runacres, M. C., \& Owocki, S. P. 2005, A\&A, 429, 323

\refer Seaquist, E. R., \& Gregory, P. C. 1973, Nature, 245, 85

\refer Smith, N., \& Owocki, S. P. 2006, ApJ, 645, L45

\refer Umana, G., Buemi, C. S., Trigilio, C., Leto, P., \& Hora, J. L.
       2010, ApJ, 718, 1036

\refer Van der Hucht, K. A., Williams, P. M., Spoelstra, T. A. Th., \&
       De Bruyn, A. C. 1992, ASP Conf. Proc., 22, 249

\refer Van Loo, S., Runacres, M. C., \& Blomme, R. 2006, A\&A, 452, 1011

\refer Van Loo, S., Blomme, R., Dougherty, S. M., \& Runacres, M. C.
       2008, A\&A, 483, 585

\refer Vink, J. S., De Koter, A., \& Lamers, H. J. G. L. M. 2000, A\&A, 362, 295

\refer Wendker, H. J., Baars, J. W. M., \& Altenhoff, W. J.
       1973, Nature, 245, 118

\refer Wendker, H. J., Smith, L. F., Israel, F. P., Habing, H. J., \&
       Dickel, H. R. 1975, A\&A, 42, 173

\refer White, R. L., \& Becker, R. H. 1983, ApJ, 272, L19

\refer White, R. L., \& Becker, R. H. 1995, ApJ, 451, 352

\refer Williams, P. M., Dougherty, S. M.; Davis, R. J., et al.
       1997, MNRAS, 289, 10

\refer Wright, A. E., \& Barlow, M. J. 1975, MNRAS, 170, 41

\endrefer           
\end{document}